\documentclass[final]{aipproc}
\usepackage{amsmath,amssymb}
\usepackage{epsf}

\layoutstyle{6x9}

\newcommand\f{\frac}
\newcommand\p{\partial}

\begin{document}

\title{Warped and eccentric discs around black holes}



\author{Gordon I. Ogilvie}{}
\author{B\'arbara T. Ferreira}
{address={Department of Applied Mathematics and Theoretical Physics,
University of Cambridge,\\
Centre for Mathematical Sciences, Wilberforce Road, Cambridge CB3 0WA, UK}}

\begin{abstract}
Accretion discs around black holes in X-ray binary stars are warped if
the spin axis of the black hole is not perpendicular to the binary
orbital plane.  They can also become eccentric through an instability
involving a resonance with the binary orbit.  Depending on the
thickness of the disc and the efficiency of dissipative processes,
these global deformations may be able to propagate into the innermost
part of the disc in the form of stationary bending or density waves.
We describe the solutions in the linear regime and discuss the
conditions under which a warp or eccentricity is likely to produce
significant activity in the inner region, which may include the
excitation of quasi-periodic oscillations.
\end{abstract}

\maketitle

\section{Introduction}

Accretion discs involve approximately Keplerian motion around a
massive central object, and the general solution allows for nested
orbits with smoothly varying inclination and eccentricity.  The shape
of a warped or eccentric disc evolves slowly under the action of
stresses in the disc and external forces that deviate from that of a
Newtonian point mass.  Coherent precession of discs is usually
possible in binary stars.

Warped and eccentric discs may be relevant to a wide variety of
phenomena in X-ray binaries.  Long-period modulations such as
superhumps and superorbital variability may be attributable to the
precession of global warping or eccentric modes that are not forced but
are excited by various instabilities.  Sufficiently large discs
encounter a resonance at the location where the angular velocity of
the disc is three times that of the binary orbit and may become
eccentric as a result \citep{W88,L91}.  This effect is well known in
cataclysmic variable stars of mass ratio $q\lesssim0.3$
\citep[e.g.][]{P05} where, as well as giving rise to superhumps, the
eccentricity enhances the viscous dissipation and significantly
affects the outburst dynamics.  Similar processes should occur in most
X-ray binaries with black hole primaries, and indeed superhumps are
reported in an increasing number of such systems
\citep{OC96,HKMC01,U02,NBC07,Z08}.  A warping instability involving
radiation forces \citep{P96} is more likely to occur with neutron star
primaries \citep{OD01} but may also be seen in GRS~1915+105, which has
a very large disc \citep{RGM03}.  Stationary warped discs also arise
whenever there is a misalignment between the orbital angular momentum
of the binary and the spin angular momentum of the central object
\citep{M02}, a knowledge of which is communicated to the disc through
general relativistic (gravitomagnetic) or magnetic torques
\citep{BP75,L99}.

The innermost parts of discs around black holes and neutron stars
depart significantly from Keplerian motion.  Indeed, the rapid
relativistic precession of elliptical or inclined orbits has often
been discussed in connection with quasi-periodic oscillations (QPOs)
in X-ray binaries, and may best be studied within the context of
eccentric or warped discs.  There is an apparent conflict between the
description of the outer part of the disc, which supports slowly
precessing global deformations, and that of the non-Keplerian inner
region.  In this paper we attempt to make this connection and to
describe how a global deformation of the disc in the form of a
stationary or slowly precessing warp or eccentricity may be able to
propagate inwards under some conditions, activating the inner region
and possibly exciting trapped oscillations that may explain
high-frequency QPOs in accreting black holes.  The excitation
mechanism itself has been studied by \citet{K03,K08} and by \citet[see
also this volume]{FO08}.

\section{Local analysis}

A small warp or eccentricity can be considered as a perturbation of a
standard (circular and coplanar) disc.  Such a disc supports a variety
of wave modes, having a dependence on time and azimuth of the form
$\exp(im\phi-i\omega t)$.  In the simplest case of a strictly
isothermal disc, a local dispersion relation
\begin{equation}
  k^2H^2=\f{(\hat\omega^2-\kappa^2)(\hat\omega^2-n\Omega_z^2)}{\hat\omega^2\Omega_z^2}
\end{equation}
can be derived \citep{OKF87}, which relates the Doppler-shifted wave
frequency $\hat\omega=\omega-m\Omega$ to the radial wavenumber $k$.
The integers $m$ and $n\ge0$ are the azimuthal and vertical mode
numbers, while $H$ is the vertical scaleheight of the disc and
$\Omega$, $\kappa$ and $\Omega_z$ are the orbital, epicyclic and
vertical oscillation frequencies characteristic of circular orbits in
the given potential or metric.  All of these quantities depend on the
radius $R$ at which the dispersion relation is evaluated.  The local
dispersion relation of a more general disc model can be calculated
numerically.  For this description to be accurate, the wavelength
$\lambda=2\pi/k$ should be much less than $R$.

Within this context, a warp corresponds to $(m,n)=(1,1)$ (vertical
motion independent of $z$) and an eccentricity to $(m,n)=(1,0)$
(horizontal motion independent of $z$).  The dispersion relation shows
that the warp takes the form of a propagating bending wave ($k^2>0$)
when $(\omega-\Omega)^2>\max(\kappa^2,\Omega_z^2)$ or
$<\min(\kappa^2,\Omega_z^2)$, while the eccentricity takes the form of
a propagating density wave when $(\omega-\Omega)^2>\kappa^2$.

\section{Secular theories}

A complementary description is provided by theories that consider a
warp or eccentricity that varies on a length-scale much longer than
$H$ and on a time-scale much longer than $\Omega^{-1}$.
Small-amplitude warps are governed by the equations
\begin{equation}
  \Sigma R^2\Omega\left[\f{\p W}{\p t}-i\left(\f{\Omega^2-\Omega_z^2}{2\Omega}\right)W\right]=\f{1}{R}\f{\p G}{\p R},
\end{equation}
\begin{equation}
  \f{\p G}{\p t}-i\left(\f{\Omega^2-\kappa^2}{2\Omega}\right)G+\alpha_W\Omega G=\f{PR^3\Omega}{4}\f{\p W}{\p R}
\end{equation}
\citep[e.g.][]{LOP02}, where $W$ describes the amplitude and phase of
the inclination of the disc at radius $R$ and time $t$, $G$ refers to
a horizontal torque communicated by Reynolds stresses, $\Sigma$ is the
surface density and $P=\Sigma H^2\Omega_z^2$ is the vertically
integrated pressure.  These equations describe propagating bending
waves with essentially the same local dispersion relation as in the
previous section.  (The theories overlap when $H\ll\lambda\ll R$,
which is possible in a thin disc.)  They also allow for viscous (i.e.\
turbulent) damping of the warp, parametrized using a dimensionless
number $\alpha_W$, which is equivalent to the usual Shakura--Sunyaev
parameter \citep{SS73} if the disc has an isotropic effective
viscosity.

The simplest equation describing a small eccentricity is
\begin{equation}
  -2i\Sigma R^2\Omega\f{\p E}{\p t}=\f{1}{R}\f{\p}{\p R}\left[(\gamma-i\alpha_E)PR^3\f{\p E}{\p R}\right]+R\f{\p P}{\p R}E+\Sigma R^2(\Omega^2-\kappa^2)E
\end{equation}
\citep[e.g.][]{GO06}, where $E$ describes the amplitude and phase of
the eccentricity of the disc at radius $R$ and time $t$, $\gamma$ is
the adiabatic exponent and $\alpha_E$ parametrizes the viscous
damping.  For short wavelengths, this equation also agrees with the
local dispersion relation of the isothermal disc when $\gamma=1$ and
$\alpha_E=0$.  This equation is based on a two-dimensional
approximation and neglects many of the complications of a shear
viscosity such as viscous overstability \citep{K78}.  More
sophisticated theories, including nonlinearity and all viscous or
viscoelastic effects, are available in the literature \citep{O00,O01}.

\section{Stationary deformations}

\begin{figure}
\label{f:lambda}
\centerline{\epsfysize=7.5cm\epsfbox{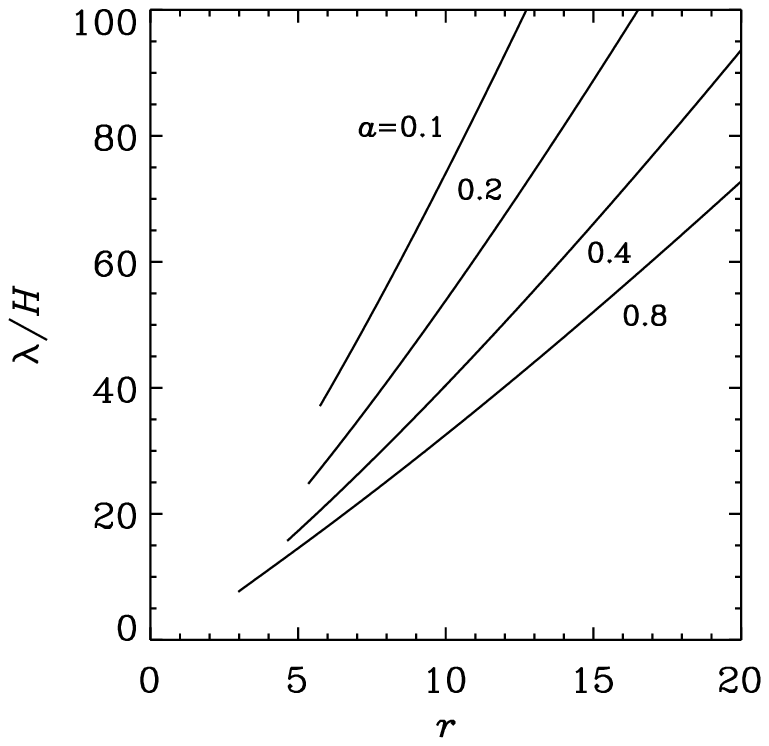}\epsfysize=7.5cm\epsfbox{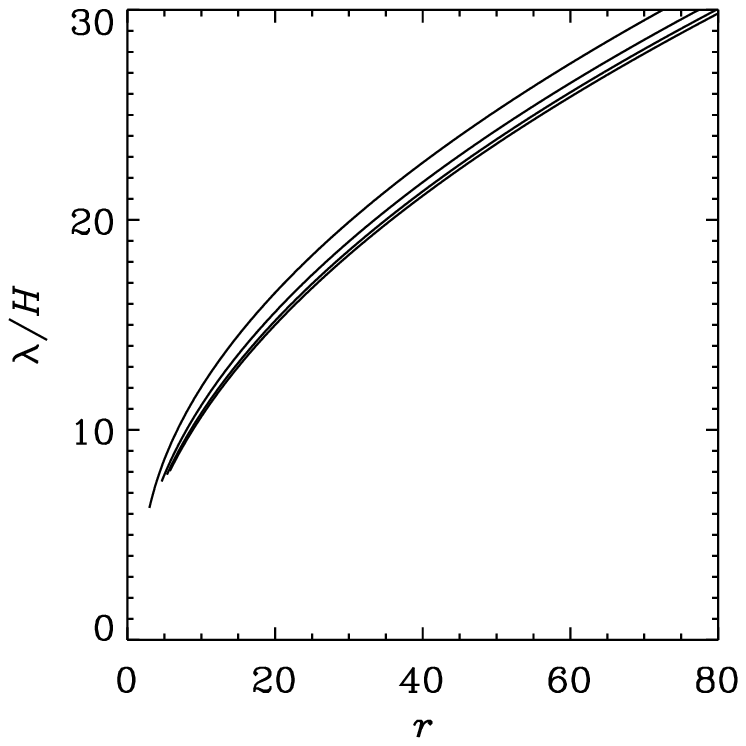}}
\caption{Local wavelength of a stationary warp (\textit{left}) or
eccentricity (\textit{right}), in units of the local vertical
scaleheight $H$, plotted versus the radius in gravitational units, for
black holes with spin parameters $a=0.1$, $0.2$, $0.4$ and $0.8$.  The
disc is terminated at the marginally stable orbit in each case.}
\end{figure}

A global warp or eccentricity precesses at only a fraction of the
binary orbital frequency, so the wave frequency $\omega$ is completely
negligible compared to $\Omega$, $\kappa$ and $\Omega_z$ in the inner
part of the disc.  Setting $\omega=0$, we find the local radial
wavenumber to be given by
$k^2H^2=(\Omega^2-\kappa^2)(\Omega^2-\Omega_z^2)/\Omega^2\Omega_z^2$
in the case of a warp and
$k^2H^2=(\Omega^2-\kappa^2)/\gamma\Omega_z^2$ for an eccentricity.
Using the expressions for $\Omega$, $\kappa$ and $\Omega_z$ for the
Kerr metric \citep{K90} and taking $\gamma=5/3$, we find $\lambda/H$
as a function of the dimensionless radius $r=Rc^2/GM$ and the spin
parameter $a$ (see Fig.~\ref{f:lambda}).  For $a>0$, both $W$ and $E$
propagate at all radii, with wavelengths everywhere significantly
longer than $H$.  Of all the wave modes described by the local
dispersion relation, these are the most credible in a turbulent disc
because of their relatively long wavelengths.

It is also possible to predict how the amplitudes of the deformations
scale with radius.  In the absence of viscous damping, a WKB analysis
of the secular theories shows that $|W|\propto R^{1/8}(\Sigma
H)^{-1/2}$ and $|E|\propto R^{1/4}(\Sigma H)^{-1/2}$.  We apply these
results to a steady accretion disc in the regime dominated by gas
pressure and Thomson opacity, in which (assuming
$\alpha=\mathrm{constant}$) $\Sigma\propto f^{3/5}R^{-3/5}$ and
$H\propto f^{1/5}R^{21/20}$, where $f=1-(R_{in}/R)^{1/2}$
\citep{SS73}.  Then $|W|\propto f^{-2/5}R^{-1/10}$ and $|E|\propto
f^{-2/5}R^{1/40}$, implying a very mild dependence of the amplitude on
radius.  However, the gradients $dW/dR$ and $dE/dR$ do increase
sharply at small $R$, because of the rapidly decreasing wavelength.

\begin{figure}
\label{f:w}
\centerline{\epsfysize=8.5cm\epsfbox{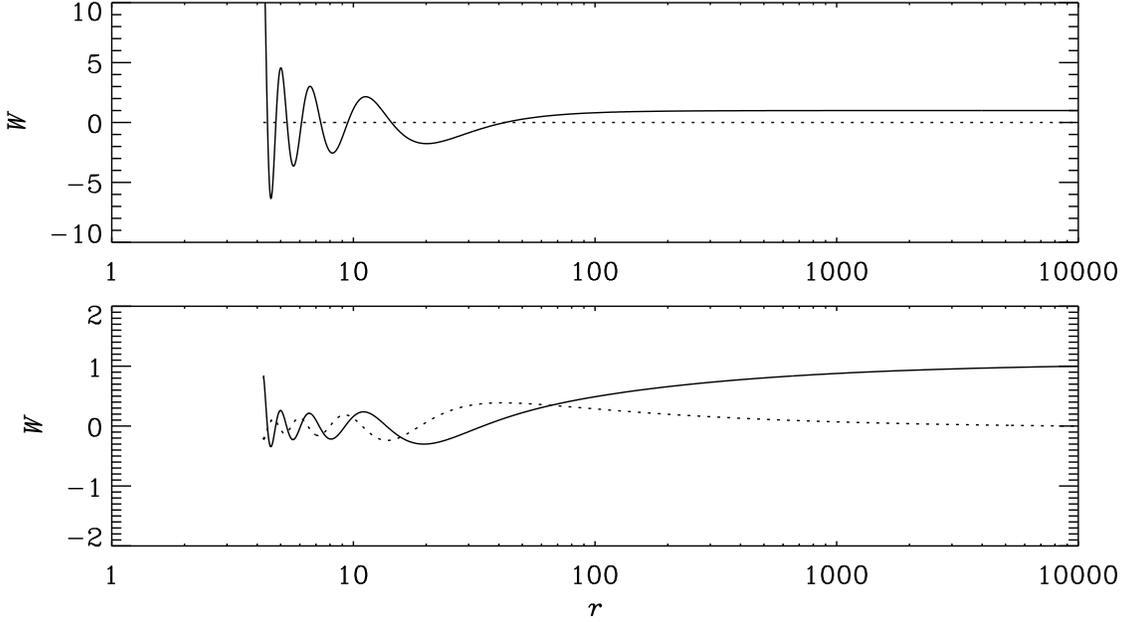}}
\caption{Stationary warp in a disc around a black hole with $a=0.5$,
$\epsilon=0.02$ and with $\alpha_W=0$ (\textit{top}) and $0.05$ (\textit{bottom}).  Real and imaginary parts of $W$ are
plotted as solid and dotted lines.  The amplitude is scaled such that $W\to1$ at large $r$.}
\end{figure}

\begin{figure}
\label{f:e}
\centerline{\epsfysize=8.5cm\epsfbox{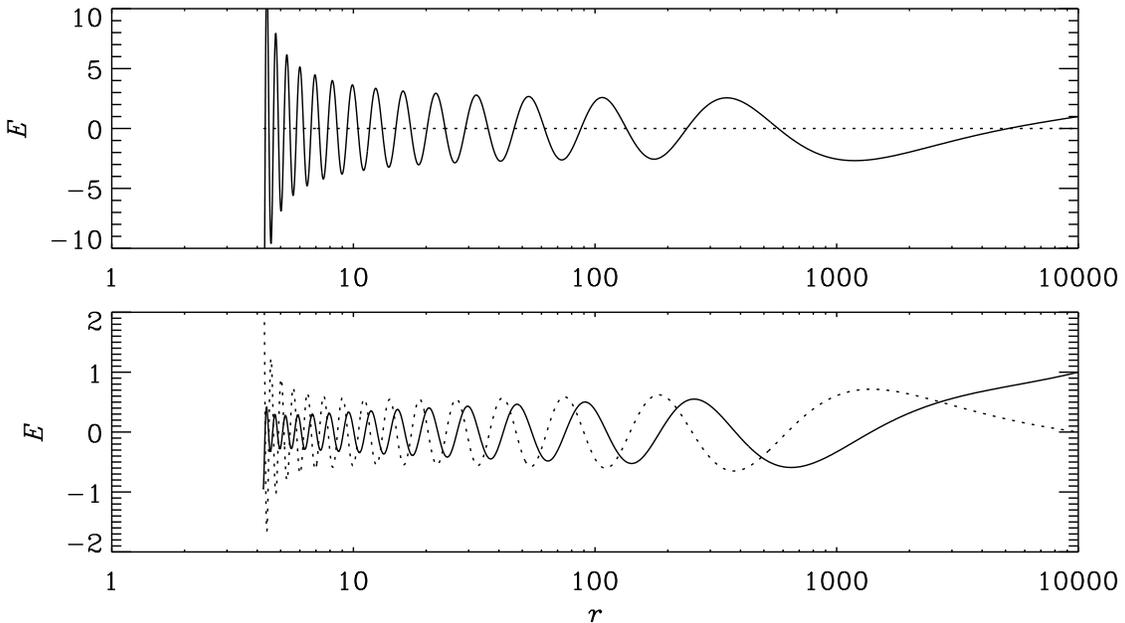}}
\caption{Stationary eccentricity in a disc around a black hole with $a=0.5$,
$\epsilon=0.02$ and with $\alpha_E=0$ (\textit{top}) and $0.05$ (\textit{bottom}).  Real and imaginary parts of $E$ are
plotted as solid and dotted lines.}
\end{figure}

When dissipation is taken into account, these solutions are modified
by viscous attenuation.  Our interest is in whether a deformation of
the outer part of the disc can propagate into the inner region with
non-negligible amplitude.  Noting that $H/R$ is almost independent of
$R$ in the above model, we define the constant parameter $\epsilon$ by
$H/R=\epsilon f^{1/5}r^{1/20}$.  Then the logarithm of the attenuation
factor for a complete crossing of the disc scales approximately with
$\alpha_W/\epsilon$ or $\alpha_E/\epsilon$.

The numerical solutions in Fig.~\ref{f:w} confirm this behaviour.  In
the absence of dissipation, the bending wave reflects perfectly from
the stress-free inner boundary and sets up a standing wave.
Significant attenuation is found when $\alpha_W$ exceeds $\epsilon$,
in which case only a wave with inward group velocity is seen.  If
$\alpha_W$ is increased much further, the oscillations are no longer
apparent.  Only in this case does the inner part of the disc lie in
the equatorial plane of the black hole as suggested by \citet{BP75}.
Note that $W$ tends to a constant at large $r$, which corresponds to
the inclination of the outer part of the disc with respect to the
black hole's equator.  The oscillatory structure of the warp has been
noted before \citep{II97,LOP02}.  Traces of it may already have been
detected in numerical simulations \citep{FBAS07}.

The behaviour of the eccentricity is similar (Fig.~\ref{f:e}) except
for the shorter wavelength (still everywhere much longer than $H$).
Again, unless $\alpha_E$ is several times greater than $\epsilon$, the
eccentricity can reach the inner region.

Several caveats accompany these solutions.  Radiation pressure, which
is more important at higher accretion rates, thickens the inner part
of the disc, increases the wavelength, and reduces the attenuation.
Viscous overstability may cause the eccentricity to grow, rather than
decay, as it propagates inwards.  Nonlinearity may be very important.
The relevant damping coefficients $\alpha_W$ and $\alpha_E$ are not
generally equal to $\alpha$ and remain relatively poorly understood.
In addition, the time for the warp or eccentricity to propagate into
the inner region and to establish the steady solutions shown here can
be long.

\section{Conclusions}

Accretion discs around black holes in X-ray binary stars may commonly
be warped or eccentric.  Depending on the thickness of the disc and
the efficiency of dissipative processes, these global deformations may
be able to propagate into the innermost part of the disc in the form
of stationary bending or density waves.  This is most likely to occur
when the disc is hotter and thicker, when the wavelengths are longer
and the viscous attenuation is less severe.  Under these conditions
the inner region may be activated and trapped oscillations may be
excited through nonlinear mode couplings \citep{K03,K08,FO08}.

\begin{theacknowledgments}
GIO acknowledges the support of STFC.  The work of BTF was supported
by FCT (Portugal) through grant no. SFRH/BD/22251/2005.
\end{theacknowledgments}

\end{document}